\documentclass[aps,prb,onecolumn,superscriptaddress,nofootinbib,floatfix,amsmath,amssymb]{revtex4-2}

\usepackage[english]{babel}

\usepackage[letterpaper,top=2cm,bottom=2cm,left=3cm,right=3cm,marginparwidth=1.75cm]{geometry}

\usepackage{amsmath}
\usepackage{graphicx}
\usepackage{mathrsfs}
\usepackage{subcaption}
\usepackage{braket}
\usepackage{hyperref}
\usepackage{caption}
\usepackage{subcaption}

\newcommand{\avg}[1]{\langle #1 \rangle}

\usepackage{caption}
\usepackage{soul}
\usepackage[utf8]{inputenc}
\usepackage{amsmath}
\usepackage{slashed}
\usepackage{graphicx}
\usepackage{xcolor}
\begin{document}
\title{The role of averages in CV-QKD over fast fading channels}
\author{Miguel Castillo-Celeita}
		\email{miguelfernando.castillo@uva.es}
		\affiliation{Departamento de Física Teórica, Atómica y Optica, Universidad de Valladolid, E-47071 Valladolid, Spain}
\author{Matteo Schiavon}
		\email{matteo.schiavon@telecom-paris.fr}
		\affiliation{Sorbonne Université - LIP6, 4 place Jussieu, 75252 Paris, France}
            \affiliation{Telecom Paris, Institut Polytechnique de Paris, F-91120 Palaiseau, France}

\begin{abstract}
    This work presents a study of continuous-variable quantum key distribution (CV-QKD) protocols over fast-fading channels, typically found in free-space communication links. Two eavesdropping models are considered to evaluate their security under collective attacks: \textit{Holevo bound average} (HBA) and \textit{covariance matrix average} (CMA). In the HBA approach, the Holevo bound is averaged over the channel transmittance. In contrast, the CMA method calculates the Holevo bound from the average covariance matrix. Analytical expressions are developed for both strategies. The two methods also differ in how they calculate the mutual information between the legitimate parties. The results demonstrate that the SKR is significantly influenced by how you treat channel fluctuations, highlighting the importance of choosing the model that better describes the actual implementation of the protocol.
\end{abstract}
\maketitle

\section{Introduction}
In recent decades, significant progress has been achieved in the field of quantum communications, enabled by the exchange of quantum states of light between different locations on Earth through either optical fiber~\cite{yin2016,huang2016,orieux2016,zhuang2025} or free-space channels~\cite{fedrizzi2009,jin2010,sidhu2021,pan2023}.
To extend communication distances, it is essential to integrate terrestrial infrastructure with space-based devices~\cite{liao2017,chen2021,xiang2023}.
However, the effective deployment of such quantum communications setups requires preventing information leakage. Thus, quantum key distribution (QKD) and continuous variable quantum key distribution (CV-QKD) methods are regarded as solutions to the security challenges that may arise in communication between two parties~\cite{diamanti2015, pirandola2020, cao2022}.
Currently, the security of CV-QKD with stationary devices is extensively studied theoretically, in contrast to free-space configurations~\cite{pirandola2021limits,zhang2024}.
In this configuration, the transmission channel is described as a stochastic process in both the amplitude and the phase of the received signal~\cite{klen2023}, due to turbulence or other environmental factors~\cite{li2021,zuo2021,hu2022,aman2023}.
Since phase fluctuations are common to the signal and the local oscillator in the transmitted local oscillator (TLO) configuration~\cite{zhang2020} and they must be corrected using pilot tones in the local local oscillator (LLO) configuration~\cite{qi2015}, their presence, despite being challenging from an experimental point of view, does not represent a particular problem from a theoretical point of view.
Amplitude fluctuations, on the other side, must be explicitly included in the security proof and have a direct impact on the secret key rate.
Two situations can be identified in the literature~\cite{papanastasiou18}: the slow fading channel, in which the transmittance is stable long enough to allow an accurate estimation, and the fast fading channel, in which the rapid variations of the channel transmittance prevent the measurement of instantaneous values. In this situation, the secret key rate depends on the statistical distribution of the transmittance.\\

Two distinct statistical approaches for collective attacks in a fast-fading channel have been identified in the literature of CV-QKD~\cite{papanastasiou18,dequal21}. The first approach~\cite{papanastasiou18} is based on the assumption that, similarly to the slow fading channel, the Holevo bound is averaged over the channel transmittance distribution, while the legitimate users are linked by a channel of minimal transmittance $T_{min}$, meaning that the shared mutual information is $I_{AB}^{T_{min}}$. For this reason, this model will be referred to as \textit{Holevo bound average} (HBA).
This model is the basis of many recent theoretical studies of free-space and satellite CV-QKD~\cite{fan2022,yang2023,zhao2024,zhao2025}. 
The alternative approach~\cite{dong2010,usenko2012} assumes that the channel is a statistical mixture of fixed transmittance channels and, due to the optimality of Gaussian attacks~\cite{navascues2006}, the state that maximizes the leaked information is the one characterized by a covariance matrix averaged over the channel transmittance distribution.
For this reason, this approach will be referred to as \textit{covariance matrix average} (CMA).
In this model, the mutual information between the legitimate parties is evaluated as $\avg{I_{AB}}$, which represents the state of the art of channel coding for this particular channel~\cite{lapidoth2002,dequal21}.

Here is presented a comparison of the fast fading transmission channel in the previous approaches, considering an idealized uniform transmission channel~\cite{papanastasiou18} that does not take into account the diffraction, pointing and tracking errors present in a real free-space or satellite channel~\cite{pirandola2021,pirandola2021satellite}.
The secret key rate is evaluated for both approaches for different uniform fluctuating channels in order to identify which method gives the most conservative result for the different configurations.
To achieve the objectives outlined above, this article is organized as follows: Section~\ref{assumptions} presents the protocol and shows the parameter estimation for a channel with fixed transmittance. Section \ref{large_variance} describes the equations for the secret key rate (SKR) for the HBA approach while Section~\ref{mean_value} develops the SKR for the CMA approach. The two methods are compared in Section~\ref{discussion}. Finally, conclusions are provided in the last section.

\section{Gaussian protocol for a fixed transmission channel}\label{assumptions}
In principle, the set of rules governing communication is established by the system parties, Alice and Bob. In this context, the Gaussian Modulated Coherent State (GMCS) CV-QKD protocol with reverse reconciliation is considered~\cite{grosshans2002,jouguet2011,weedbrook2012}.
While this is a prepare-and-measure protocol, its theoretical study is easier considering the equivalent entanglement-based protocol that we will describe here.

\begin{figure}
    \centering
    \includegraphics[width=0.7\linewidth]{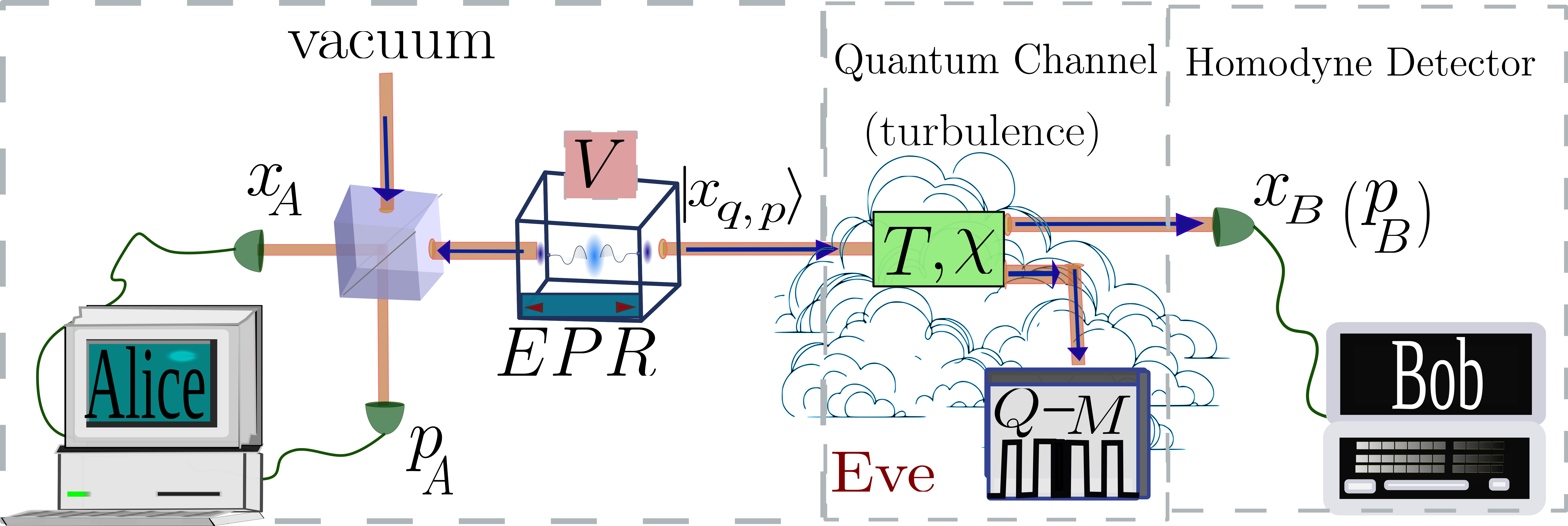}
    \caption{This is the scheme of the system, where Alice prepares a two-mode coherent state and sends half through a quantum channel with transmittance $T$ and noise variance $\chi$. Eve can interact with this mode by injecting extra noise and storing the state information in a quantum memory (Q-M). Bob receives the output mode and measures one of the quadratures in a homodyne detector.}
    \label{fig:scheme}
\end{figure}
The protocol, whose scheme is shown in Figure~\ref{fig:scheme}, can be modeled with the following steps:
\begin{itemize}
    \item Alice prepares a two-mode squeezed vacuum (TMSV) state of variance $V$ and measures one mode using a heterodyne detector.
    This projects the other mode $B$ into the coherent state $\ket{x_{q,p}}$, whose average quadratures $\left( \overline{p}, \overline{q} \right)$ are distributed according to a zero-centered Gaussian distribution of variance $V_A = V - 1$.
    \item Alice transmits the mode $B$ through a quantum channel with transmittance $T$ and noise $\chi$.
    \item Bob receives the mode $B$ and applies homodyne detection on one of the quadratures $q$ or $p$, randomly chosen with uniform distribution.

\end{itemize}
The quadratures of the state received by Bob have a variance which is affected by the transmittance $T$ and the channel noise $\chi$,
\begin{equation}
    V_B=T(V+\chi), \qquad \chi=\frac{1}{T}-1+\varepsilon,
    \label{eq:01}
\end{equation}
where the quantities are normalized to shot-noise units (SNU) and $\varepsilon$ is the excess noise referred to channel input.
In this work, we chose to include detector efficiency and electronic noise into the channel parameters.
While this gives worse results in terms of SKR, it allows us to directly compare our results with the relevant literature~\cite{papanastasiou18}.\\

The SKR in the case of collective attacks is evaluated using the formula~\cite{scarani2009}
\begin{equation}
    R = I_{AB} - I_E,
    \label{eq:02}
\end{equation}
where $I_{AB}$ is the mutual information shared between Alice and Bob and $I_E$ represents the information leaked to the eavesdropper Eve.
Since the random variables obtained by Alice and Bob after the measurement are Gaussian, the mutual information is derived as the difference between the differential entropy~\footnote{The differential entropy, a continuous version of the Shannon entropy, has the form $\mathcal{H}(X)=-\int dx\mathcal{P}(x)\log_2\mathcal{P}(x)$, which is equal to $1/2 \log_2(2 \pi e) + 1/2 \log_2 V$ for a Gaussian random variable of variance $V$.} of Bob $\mathcal{H}(X_B)$ and the differential entropy of Bob conditioned on Alice's variable $\mathcal{H}(X_{B|A})$, with $V_{B|A}=T(1+\chi)$~\cite{fossier09}.
Consequently, the mutual information is given by 
\begin{equation}
    I_{AB}=\frac{1}{2}\log_2\left(\frac{V_B}{V_{B|A}}\right)=\frac{1}{2}\log_2\left(\frac{V+\chi}{1+\chi}\right).
    \label{eq:03}
\end{equation}

\subsection{Information leaked to Eve under collective attacks}\label{fossier}
In this work, it is assumed that the eavesdropper implements a collective attack, which is equivalent to the most general possible quantum attack in the asymptotic limit~\cite{leverrier2010}. In this scenario, Eve interacts separately with each mode transmitted through the channel using ancillary modes stored in a quantum memory, which are then measured jointly at her will.

In the above scenario, the leaked information is bounded by the Holevo bound
\begin{equation}
    I_E \equiv \chi_{\text{BE}}=S(E) - S(E|B) =S(\rho_{E})-\int dm_B\,p(m_B)\,S(\rho^{m_B}_{E}),
    \label{eq:04}
\end{equation} 
where $S(\cdot)$ is the Von Neumann entropy and $m_B$ represents the result of Bob measurement. Moreover, under the assumption that Eve owns a purification of the system, the Holevo quantity admits the following expression
\begin{equation}
        \chi_{BE}=S(E)-S(E|B)=S(AB)-S(A|B).
        \label{eq:05}
\end{equation}
The optimality of Gaussian attacks enables the computation of the entropies as~\cite{serafini17}
\begin{equation}
    \chi_{BE}=\sum_{i=1}^2G\left(\frac{\lambda_i-1}{2}\right)-G\left(\frac{\lambda_3-1}{2}\right),
    \label{eq:06}
\end{equation}
where $G(x)=(x+1)\log_2(x+1)-x\log_2x$.
In this case, the $\lambda_{1,2}$ represent the symplectic eigenvalues of the covariance matrix $\gamma_{AB}$, described the state jointly shared between Alice and Bob, while $\lambda_3$ represents the symplectic eigenvalue of the covariance matrix $\gamma_{A|B}$, which corresponds to the conditional covariance matrix after Bob measurement.
In this manner, the covariance matrix $\gamma_{AB}$, corresponding to the bipartite state $\rho_{AB}$, can be expressed as
\begin{equation}
   \gamma_{AB}=
   \left(\begin{matrix} \gamma_A & \sigma_{AB}^T \\
   \sigma_{AB} & \gamma_B \end{matrix}\right) =
   \left(\begin{matrix}
V\,\textbf{1} & \sqrt{T(V^2-1)}\sigma_z\\
\sqrt{T(V^2-1)}\sigma_z & T(V+\chi)\,\textbf{1} 
\end{matrix}\right).
\label{eq:07}
\end{equation}
The symplectic eigenvalues correspond to the eigenvalues of the matrix $\tilde{\gamma}_{AB}$ defined as
\begin{equation}
  \tilde{\gamma}_{AB}=i\Omega\,\gamma_{AB}, \quad \Omega=\oplus_{l=1}^2\left(\begin{matrix}0 & 1\\ -1 & 0\end{matrix}\right).
  \label{eq:08}
\end{equation}
They are given by
\begin{equation}
    \lambda_{1,2}=\sqrt{\frac{1}{2} \left(A\pm\sqrt{A^2-4 B}\right)},
    \label{eq:09}
\end{equation}
where
\begin{equation}
    A=T^2 (V+\chi)^2+(1-2 T) V^2+2 T, \quad B=T^2 (V \chi+1)^2,
    \label{eq:10}
\end{equation}
which may equivalently be expressed as
\begin{equation*}
    A=\left(A_0^2+(1-2 T)+\frac{2T}{V^2}\right)V^2, \quad B=B_0^2\,V^2.
    \label{eq:10b}
\end{equation*}
with $A_0=T\left(1+\chi/V\right)$ and $B_0=T\left(\chi+1/V\right)$.
The second term of Eq.~\eqref{eq:05} is calculated from the state $\rho_{A|B}$, whose covariance matrix can be written as~\cite{laudenbach18}
\begin{equation}
    \gamma_{A|B}=\gamma_A- \sigma_{AB}^T H_{\text{hom}}\sigma_{AB},
     \label{eq:11}
\end{equation}
where $H_{\text{hom}}=(X\,\gamma_{B}\,X)^{-1}$ is the symplectic matrix that represents the homodyne measurement of the $X$ quadrature. In this case, the inverse corresponds to the Moore-Penrose pseudoinverse, which, for a diagonal matrix, is obtained by inverting the elements along the diagonal $\text{diag}(a_1,\dots a_n)^{-1}=\text{diag}(a_1^{-1},\dots a_n^{-1})$.
With $X=\text{diag}(1,0)$ and $\gamma_{B}=T(V+\chi)\textbf{1}$, the matrix is expressed as
\begin{align}
   H_{\text{hom}}=\left(\begin{matrix}T(V+\chi) & 0 \\ 0 & 0\end{matrix}\right)^{-1}\nonumber=\left(\begin{matrix}\frac{1}{T(V+\chi)} & 0 \\ 0 & 0\end{matrix}\right).
   \label{eq:11a}
\end{align}
Developing the second term on the right-hand side of Eq.~\eqref{eq:11} it is possible to obtain
\begin{equation}
    \sigma_{AB}^TH_{\text{hom}}\sigma_{AB}=\frac{T(V^2-1)}{T(V+\chi)}\left(\begin{matrix} 1 & 0 \\ 0 & 0\end{matrix}\right).
    \label{eq:12}
\end{equation}
The explicit covariance matrix conditioned on the homodyne measurement is given by
\begin{equation}
    \gamma_{A|B}=V\left(\begin{matrix} 1&0\\0&1\end{matrix}\right)-\frac{V^2-1}{V+\chi}\left(\begin{matrix} 1 & 0 \\ 0 & 0\end{matrix}\right),
\label{eq:13}
\end{equation}
whose sympletic eigenvalue is
\begin{equation}
    \lambda_3=\sqrt{\frac{V(1+V\chi)}{V+\chi}}=\sqrt{\frac{B_0}{A_0}}V^{1/2}.
    \label{eq:14}
\end{equation}
A similar calculation shows that the same eigenvalue $\lambda_3$ is also obtained for a measurement of the quadrature $P$.
\section{Secret key rate with Holevo bound average}\label{large_variance}
The fast fading channel is characterized by the fact that the users can only estimate the probability distribution of the transmission efficiency and not its instantaneous value~\cite{papanastasiou18}.
An approach to deal with this problem consists in having the two legitimate parties choose the worst value for the parameters compatible with the observed distribution, using a code of rate $I_{AB}^{T_{min}}$.
Since the distinction between slow and fast fading is irrelevant for the all powerful Eve, it is possible to assume that the leaked information is the same for the two cases~\cite{papanastasiou18}.
These two assumptions represent the core of the HBA approach, which gives a SKR
\begin{align} 
    R=I_{AB}^{T_{min}}-\avg{I_{BE}},
    \label{eq:15}
\end{align}
where the average is taken on the probability distribution of the transmission efficiency.
Assuming $T \in [T_{min}, T_{max}]$ with uniform distribution and $T_{max}=T_{min}+\Delta T$, the SKR becomes
\begin{align}
R(T_{min})=I_{AB}^{T_{min}}-\frac{1}{\Delta T}\int_{T_{min}}^{T_{max}}\chi_{BE}(T)\,dT.
    \label{eq:16}
\end{align}
This approach shows an interesting behaviour in a high variance regime $V\gg 1$. In this context, the approximation for the mutual information~\eqref{eq:03} is given by
\begin{equation}
    I_{AB}^T=\frac{1}{2}\log_2 \left(\frac{T}{T+(1-T)\,\omega}\right)+\frac{1}{2}\log_2 V,
     \label{eq:17}
\end{equation}
where the $\omega$ is the thermal variance~\cite{garcia2007},
\begin{equation}
\omega=1+\frac{T\,\varepsilon}{1-T},\qquad\chi=\frac{(1-T)}{T}\,\omega.
 \label{eq:18}
\end{equation}
Before computing the Holevo bound~\eqref{eq:06}, it is necessary to approximate the expression for the eigenvalues $\lambda_{1,2,3}$ (see Appendix~\ref{appendix0} for deduction)
\begin{equation}
    \lambda_{1}\approx V(1-T),\qquad  \lambda_2\approx\omega.
     \label{eq:19}
\end{equation}
Finally, the approximation of $\lambda_3$ leads to the expression 
\begin{equation}
    \lambda_3\approx\sqrt{\frac{(1-T)\,\omega\, V}{T}}.
     \label{eq:21}
\end{equation}
Therefore, the Holevo bound $\chi_{BE}$ equation is reorganized as
\begin{align}
    \chi_{BE}(T)= & \frac{1}{2}\log_2\left(\frac{T\,(1-T)\,V}{\omega}\right)+h\left(\frac{\omega-1}{2}\right),
     \label{eq:22}
\end{align}
where $h(\omega)=(\omega+1)\log_2(\omega+1)-\omega\log_2(\omega)$.\\
In this manner, the average Holevo bound can be written in  an analytic form as

\begin{align}
    \avg{I_{BE}}=&\frac{T_{min}}{2\Delta T}\log _2\left(\frac{T_{max} \overline{T}_{max}\,\omega_{min}}{T_{min} \overline{T}_{min}\,\omega_{max}}\right)+\frac{ (1-\varepsilon)^{-1}}{2\Delta T}\log_2\left(\frac{\overline{T}_{max}\,\omega_{max}}{\overline{T}_{min}\,\omega_{min}}\right)\nonumber\\
          &-\log_2\left(\frac{T_{max}\overline{T}_{max}}{\omega_{max}}\right)+\frac{1}{\Delta T}\log_2\left(\frac{\overline{T}_{max}}{\overline{T}_{min}}\right)-\log _2(e)+\frac{1}{2}\log_2 V+\widetilde{h},
      \label{eq:23}
\end{align}
where $\overline{T}=1-T$, $\omega_{min,max}=\omega(T_{min,max})$, and the function $\widetilde{h}$ is defined in Appendix~\ref{appendix1} and is independent of $V$.
It is worth noting that, in the asymptotic regime $V\gg1$, the SKR given in Eq.~\eqref{eq:16} is independent of the variance $V$.

\section{Analysis of the SKR in the covariance matrix average approach}\label{mean_value}
The approach defined in the previous section assumes that the legitimate parties adjust the rate of the code to the minimum transmittance compatible with the channel parameters.
However, this is not the most efficient way to proceed in the case of a fluctuating channel.
The channel outputs are modeled as $Y_k=\sqrt{T_k}\,X_k+Z_k$, for the $k$-th channel use, where $T_k\in[0,1]$ denotes the ergodic fading and $Z_k\sim {\cal N}(0,\sigma)$ ergodic Gaussian noise of variance $\sigma$.\\
When fading is unknown, the receiver maps a sequence $\{y_i\}$ of received messages from a block of length $N$. This may lead to an overestimation of the channel capacity. To moderate this, the receiver incorporates noise information by mapping the pairs $(y_i,\tau_i)$, which provides a more accurate representation of a free-space communication channel~\cite{lapidoth2002}.
In this context, the receiver strategy must account for the random fluctuations introduced by fading, which otherwise may lead to an overestimation of the channel performance. To address this, the concept of ergodic capacity becomes a natural tool, as it captures the long-term average behavior of the channel under varying transmittance. By averaging over all possible realizations of the fading process, the ergodic capacity provides an appropriate characterization of the channel transmittance. Accordingly, the average mutual information shared between Alice and Bob is given by
\begin{equation}
    \langle I_{AB} \rangle=\frac{1}{\Delta T}\int_{T_{min}}^{T_{min}+\Delta T} \frac{1}{2}\log_2\left(1+\frac{T\,V_A}{\sigma}\right)\,dT,
    \label{eq:24}
\end{equation}
where the denominator is defined as $\sigma=1+\varepsilon\,T$. As a result, the mutual information between can be written in closed form as follows
\begin{align}
   \langle I_{AB} \rangle= &\frac{1}{2\,\Delta T}\bigg[ \frac{1}{\varepsilon }\log_2 \left(\frac{1+\varepsilon \,T_{min}}{1+\varepsilon  T_{max}}\right)+T_{max} \log_2 \left(\frac{1+T_{max} (\varepsilon +V_A)}{1+\varepsilon\,T_{max}}\right)\nonumber\\
    &+\frac{1}{\varepsilon +V_A}\log_2 \left(\frac{1+T_{max} (\varepsilon +V_A)}{1+T_{min} (\varepsilon +V_A)}\right)+T_{min}\log_2 \left(\frac{1+\varepsilon  T_{min}}{1+T_{min} (\varepsilon +V_A)}\right)\bigg].
    \label{eq:25}
\end{align}

The calculation of the Holevo bound in the CMA approach is based on the idea that, if the statistics over channels with transmittance fluctuations are characterized by a probability distribution $P(T)$, the channel is described by a classical mixture of different subchannels, each of which has approximately constant transmissivity $T_i$ with probability $p_i$.
Therefore $\sum_i p_i=1$, or in the continuum $\int_0^{T_{max}}p(T)\,dT=1$.
Given a Gaussian input state $\rho_{in}$, the output state of each subchannel is the Gaussian state $\rho_i$ and the final mix state is the generally non-Gaussian state $\rho=\sum_i p_i\,\rho_i$\footnote{Even if the output state is non-Gaussian, the optimality of the Gaussian attack~\cite{navascues2006} allows the use of the Gaussian state with identical moments for the calculation of the Holevo bound.}.
Similarly, the Wigner function of the output state is constructed as a weighted sum of the Wigner functions corresponding to each subchannel $W(\overline{q},\overline{p})=\Sigma_i\,p_iW_i(\overline{q},\overline{p})$, with probabilistic weights $p_i$~\cite{walschaers2021}. In this manner, the resulting function allows for the calculation of the second-order moments of the quadrature variables $(\overline{q},\overline{p})$. The covariance matrix of the output state is obtained as the average of the covariance matrices associated with each subchannel over the probability distribution $P(T)$~\cite{hosseinidehaj2021}
\begin{equation}
   \langle\gamma_{AB}\rangle=\left(\begin{matrix}
V\,\textbf{1} & \langle\sqrt{T}\rangle\sqrt{(V^2-1)}\sigma_z\\
\langle\sqrt{T}\rangle\sqrt{(V^2-1)}\sigma_z & \langle T \rangle(V+\langle\chi\rangle)\,\textbf{1} 
\end{matrix}\right).
\label{eq:26}
\end{equation}
The average channel noise is given by $ \langle\chi\rangle=1/\langle T\rangle-1+\varepsilon$, assuming that the excess noise $\varepsilon$ is independent on the channel transmittance. 
Using this expression for the noise, we get the result presented in~\cite{dequal21}
\begin{equation}
   \langle\gamma_{AB}\rangle=\left(\begin{matrix}
V\,\textbf{1} & \langle\sqrt{T}\rangle\sqrt{(V^2-1)}\sigma_z\\
\langle\sqrt{T}\rangle\sqrt{(V^2-1)}\sigma_z & (\langle T \rangle(V_A+\varepsilon)+1)\,\textbf{1} 
\end{matrix}\right).
\label{eq:27}
\end{equation}
The average conditional covariance matrix is derived in the same way of~\eqref{eq:11}
\begin{equation}
    \langle\gamma_{A|B}\rangle=V\textbf{1}-\text{diag}(1,0)\frac{\avg{\sqrt{T}}(V^2-1)}{\avg{T}(V+\langle\chi\rangle)}.
    \label{eq:28}
\end{equation}
The average covariance matrix is dependent on the two moments $\langle T \rangle$ and $\langle \sqrt{T} \rangle$ of the probability distribution of the transmittance efficiency $P(T)$.
Assuming $T \in [T_{min}, T_{max}]$ with uniform distribution, these moments can be calculated analytically as 
\begin{align}
    \langle\sqrt{T}\rangle=&\frac{1}{\Delta T}\int_{T_{min}}^{T_{min}+\Delta T}\sqrt{T}\,dT=\frac{2}{3\,\Delta T}((T_{min}+\Delta T)^{3/2}-T_{min}^{3/2}),\nonumber\\
    & \langle T\rangle=\frac{1}{\Delta T}\int_{T_{min}}^{T_{min}+\Delta T}T\,dT=T_{min}+\frac{\Delta T}{2}.
    \label{eq:29}
\end{align}
By defining the effective values of the channel parameters~\cite{vidarte2019}
\begin{align}
    & T_{eff} =\avg{\sqrt{T}}^2,\nonumber\\
    & \chi_{eff}=\frac{1}{T_{eff}}-1+\varepsilon_{eff},\nonumber\\
    & \varepsilon_{eff}=\varepsilon\left(1+\frac{\text{Var}(\sqrt{T})}{T_{eff}}\right)+\frac{\text{Var}(\sqrt{T})}{T_{eff}}V_A,
    \label{eq:30}
\end{align}
where the variance is defined by $\text{Var}(\sqrt{T}) = \avg{(\sqrt{T})^2} -\avg{\sqrt{T}}^2$ and $V_A=V-1$, it is possible to write the average covariance matrices $\langle \gamma_{AB} \rangle$ and $\langle \gamma_{A|B} \rangle$ in the same form as the fixed transmission channel ones defined in Section~\ref{fossier} by making the substitution $T \rightarrow T_{eff}$ and $\chi \rightarrow \chi_{eff}$.

Therefore, it is possible to use the same equations derived in that section for the calculation of the symplectic eigenvalues contributing to the Holevo bound~\eqref{eq:06}. Thus, the eigenvalues of the covariance matrix~\eqref{eq:27} are given as follows
\begin{equation}
    \widetilde{\lambda}_{1,2}=\sqrt{\frac{1}{2}}\sqrt{A\pm\sqrt{A^2-4B}}
    \label{eq:31}
\end{equation}
where
\begin{align}
    A=V^2(1-&2\,T_{eff})+2\,T_{eff}+T_{eff}^2(V+\chi_{eff})^2,\nonumber\\
    &B=T_{eff}^2\left(V\,\chi_{eff}+1 \right)^2.
    \label{eq:32}
\end{align}
The last eigenvalue is given by
\begin{equation}
    \widetilde{\lambda}_3=\sqrt{V \left(\frac{1+V\chi_{eff}}{V+ \chi_{eff}}\right)}.
    \label{eq:33}
\end{equation}
Using these eigenvalues, it is possible to calculate the secret key rate as
\begin{equation}
    R=\avg{I_{AB}}-\chi_{BE}(\widetilde{\lambda}_1, \widetilde{\lambda}_2, \widetilde{\lambda}_3),
    \label{eq:34}
\end{equation}
Where the Holevo bound is given by~\ref{eq:06} with the respective considerations on the transmittance distribution.

\section{Results and Discussion}
\label{discussion}
This section compares the two approaches, focusing on their influence on the SKR.
We assume for the channel transmittance an uniform distribution in the interval $[T_{min}, T_{max}]$ and define $\Delta T := T_{max} - T_{min}$, while the excess noise $\varepsilon$ is independent of the transmittance.
The results show that the achievable SKR is strongly affected by the statistical treatment of the Holevo bound, $\chi_{BE}$, during the key extraction process. Although the mutual information $I_{AB}$ plays a role, its influence is relatively minor under the conditions considered.

\begin{figure}[h!]
   \centering
    \begin{subfigure}[b]{0.48\textwidth}
         \centering
         \includegraphics[width=\textwidth]{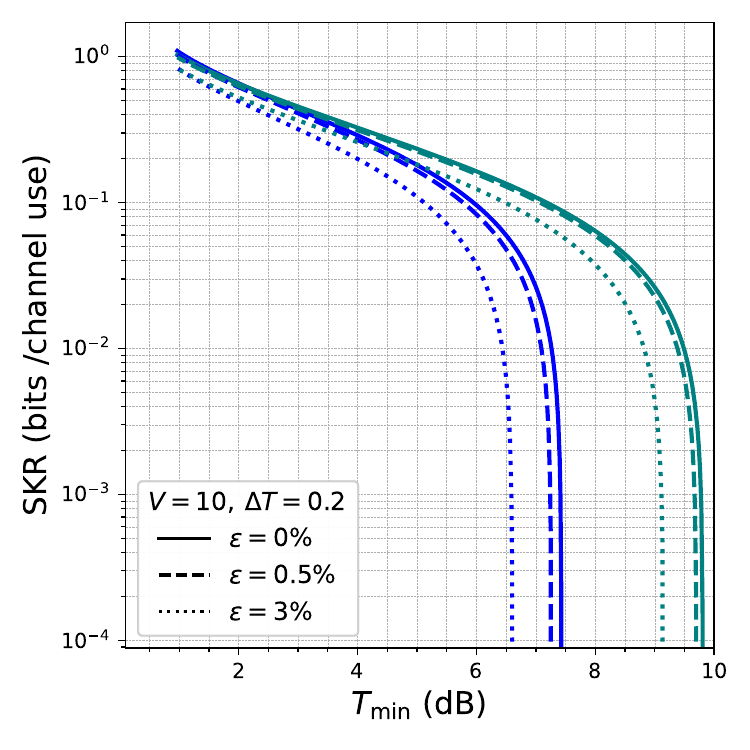}
     \end{subfigure}
         \centering
   \begin{subfigure}[b]{0.48\textwidth}
         \centering
         \includegraphics[width=\textwidth]{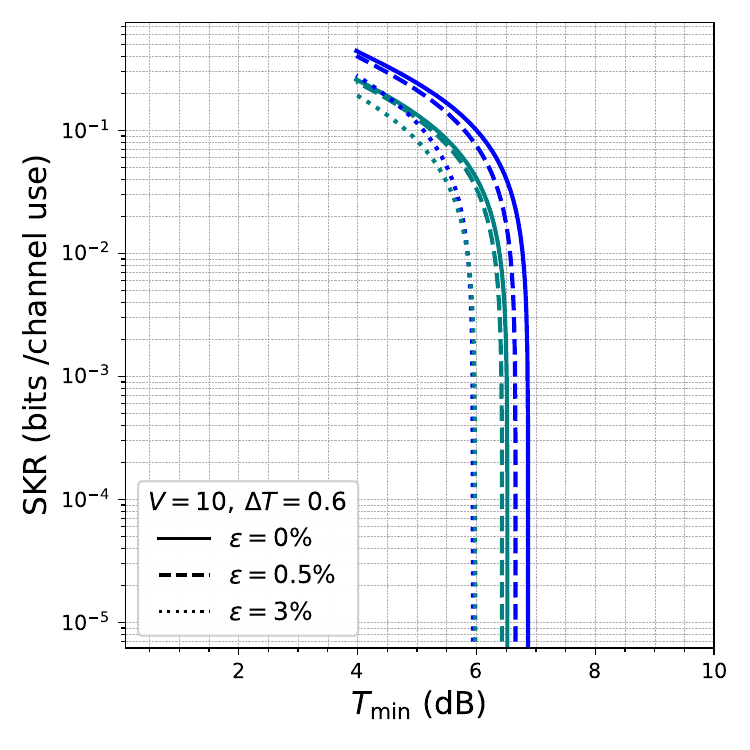}
     \end{subfigure}
     \centering
    \begin{subfigure}[b]{0.48\textwidth}
         \centering
         \includegraphics[width=\textwidth]{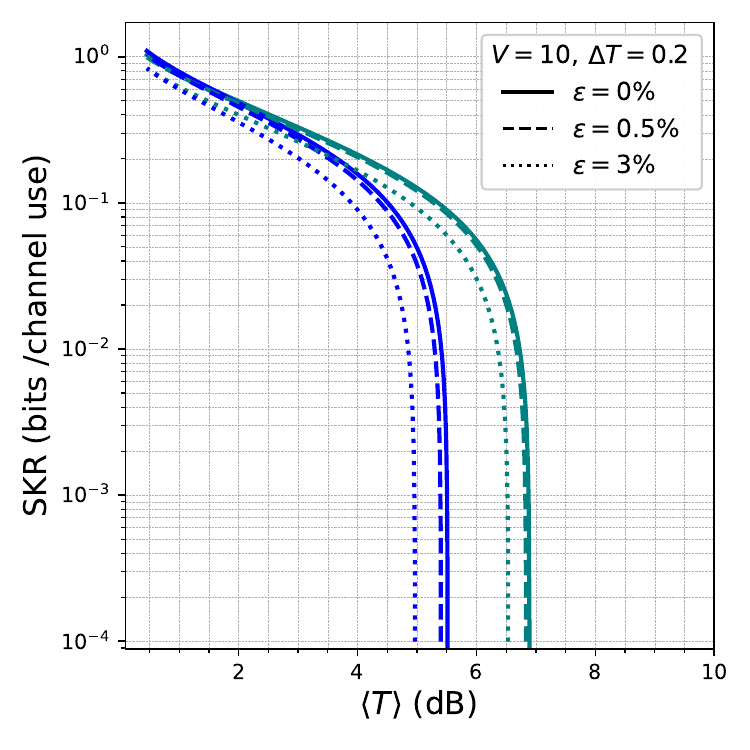}
     \end{subfigure}
         \centering
   \begin{subfigure}[b]{0.48\textwidth}
         \centering
         \includegraphics[width=\textwidth]{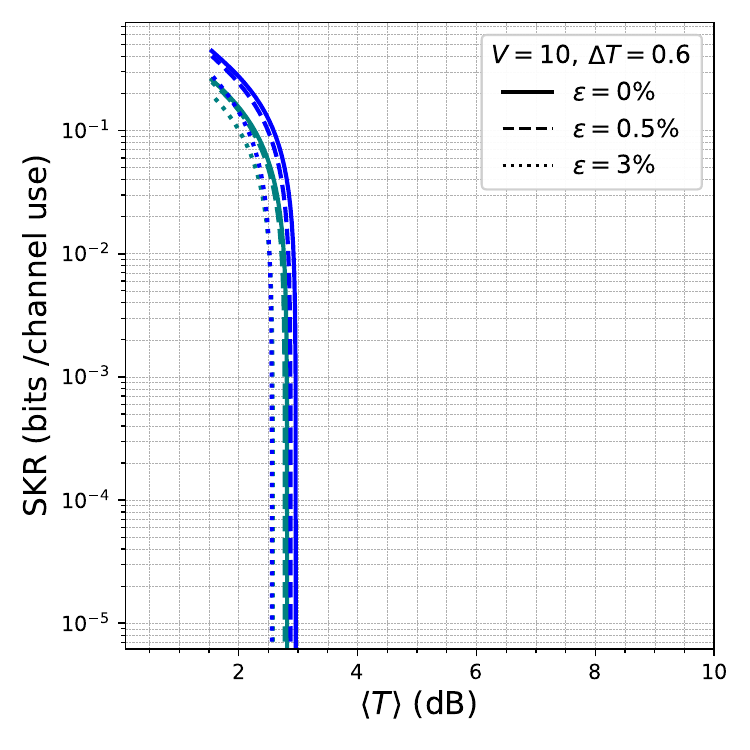}
     \end{subfigure}
 \caption{This graph presents four panels that evaluate the performance of two distinct approaches in the low-variance regime, $V = 10$. The HBA approach is shown in blue, while the CMA approach is shown in green. The figure compares the SKR for both approaches under three different excess noise levels: $0\%$ (solid lines), $0.5\%$ (dashed lines), and $3\%$ (dotted lines). Results are displayed for two transmittance ranges: $\Delta T = 0.2$ (left panels) and $\Delta T = 0.6$ (right panels). The top row displays the SKR as a function of the minimum transmittance value and the bottom row as a function of the mean transmittance.}
\label{fig:02}
\end{figure}

To evaluate the performance of both approaches under different fading regimes, the SKR behavior is evaluated across various parameter configurations ($V$, $\Delta T$, and $\varepsilon$). In Fig.~\ref{fig:02}, it is shown that, for low modulation variance, the HBA approach (blue lines) allows the generation of a positive secret key down to a $T_{min}$ of $6$–$8\,\mathrm{dB}$, regardless of the width of the transmittance distribution $\Delta T$. This result shows a relative stability of the HBA approach against channel fluctuations, which do not influence the mutual information between the legitimate parties and have a minor effect on the average Holevo bound.\\
In contrast, the CMA approach (Fig.~\ref{fig:02}, green lines) exhibits a notable degradation in the SKR as the parameter $\Delta T$ increases from $0.2$ to $0.6$. This behavior is attributed to the reduced average mutual information shared by the legitimate parties.
For both approaches, an increase in excess noise $\varepsilon$ leads to a reduction in SKR, a result that aligns with theoretical expectations and previous analyses. This trend persists across the whole range of variances and transmittance intervals.\\

\begin{figure}[h]
    \centering
      \centering
    \begin{subfigure}[b]{0.48\textwidth}
         \centering
         \includegraphics[width=\textwidth]{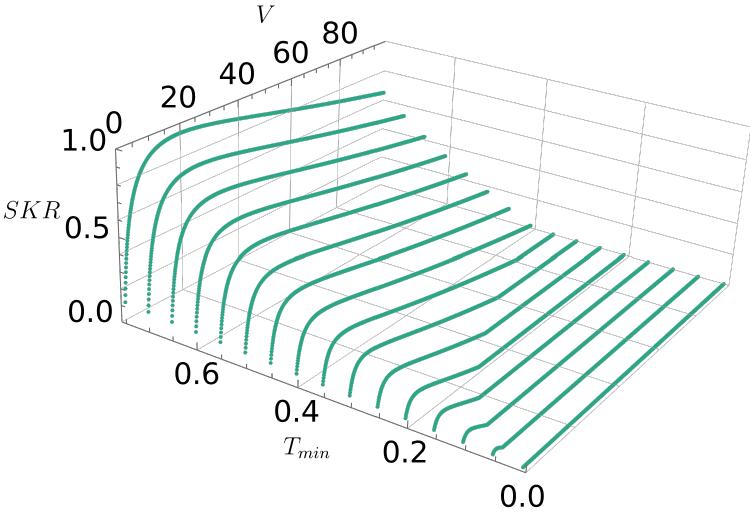}
     \end{subfigure}
         \centering
   \begin{subfigure}[b]{0.48\textwidth}
         \centering
         \includegraphics[width=\textwidth]{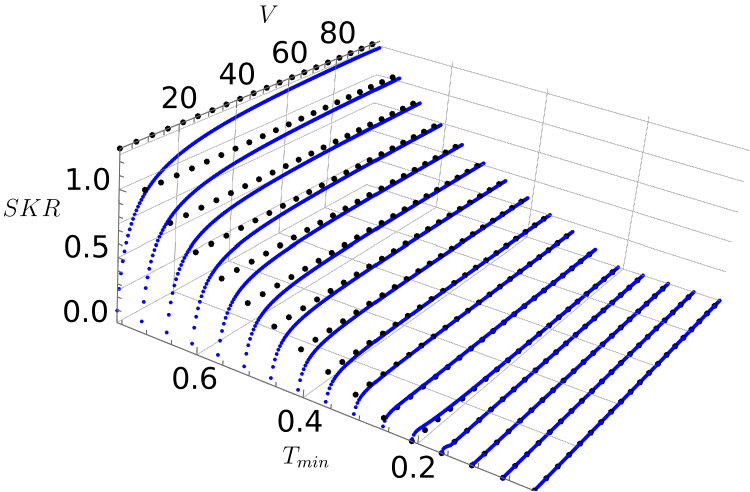}
     \end{subfigure}
    \caption{Secret key rate as a function of the modulation variance for different values of $T_{min}$. (right) The blue lines represent the HBA approach, while the horizontal black lines correspond to the SKR calculated in the high variance regime $V\gg 1$. This confirms that the optimal SKR is obtained for $V\rightarrow\infty$. (left) The green lines show the CMA approach. In this case, the optimal variance varies depending on the transmittance value and must be reduced as $T_{min}$ approaches $0$. The SKR is computed considering $\Delta T=0.2$. }
    \label{fig:03}
\end{figure}

\begin{figure}[h]
   \centering
    \begin{subfigure}[b]{0.49\textwidth}
         \centering
         \includegraphics[width=\textwidth]{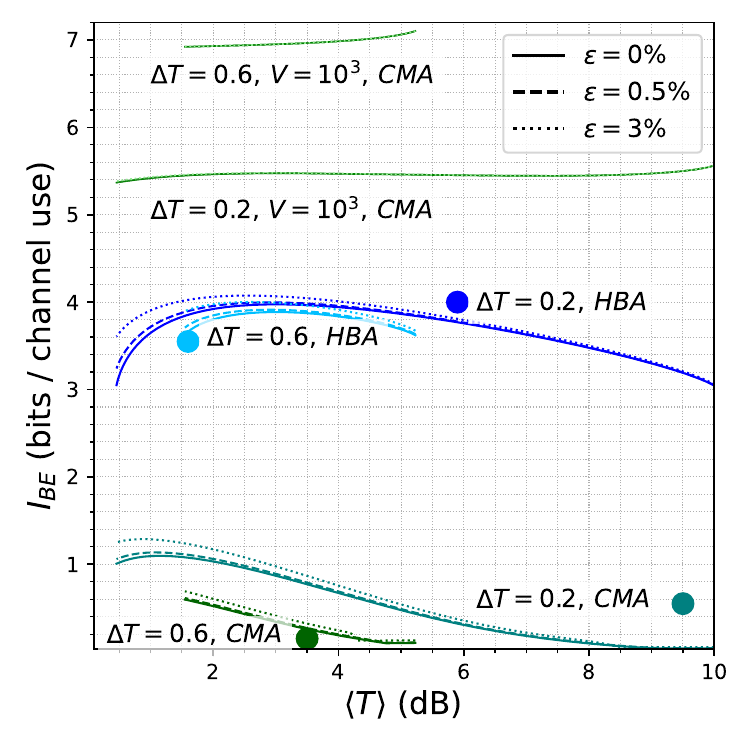}
     \end{subfigure}
         \centering
   \begin{subfigure}[b]{0.49\textwidth}
         \centering
         \includegraphics[width=\textwidth]{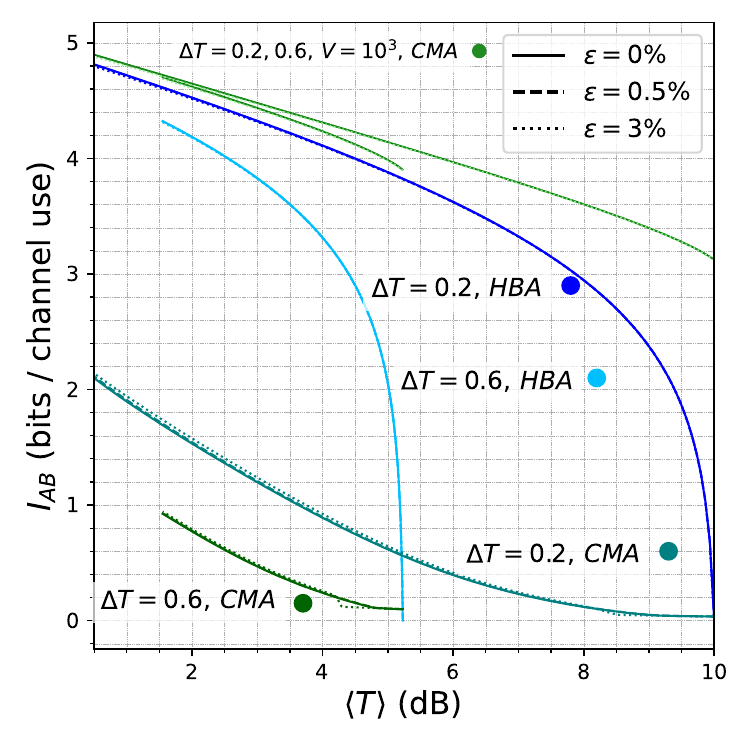}
     \end{subfigure}
 \caption{(Left) Holevo bound for the HBA (Eq.~\eqref{eq:23}), plotted in blue with variance fixed at $V=10^3$. CMA results are shown in green: the lower curves correspond to the optimal variance $V=V_{\mathrm{opt}}$ (Eq.~\eqref{eq:34}), while the upper curves use $V=10^3$. (Right) Mutual information for the HBA (Eq.~\eqref{eq:17}) and CMA (Eq.~\eqref{eq:25}), using the same color scheme (blue for HBA; green for CMA).}
\label{fig:05}
\end{figure}

Fig.~\ref{fig:03} further explores how the SKR depends on the modulation variance $V$. For the HBA approach (blue lines), the SKR increases with $V$ and eventually reaches an asymptotic regime for $V \gg 1$ (horizontal black lines), in line with the analytical result found in Section~\ref{large_variance}.
Conversely, in the CMA approach, as evident from Fig.~\ref{fig:03}, the SKR is higher in the low-variance regime but suffers a rapid decline as $V$ increases.
This effect is particularly strong for high attenuation channels and reflects the stronger influence of the modulation variance on the information leaked to the adversary than on the information shared by the legitimate parties.
Therefore, an important distinction between the two approaches in terms of the variance $V$ lies in how they scale with system variance $V$. This is rooted in the structural differences in the components that determine the symplectic eigenvalues of the Holevo bound. Specifically, in the CMA approach, the factor $B$ scales as $V^4$ (Eq.\eqref{eq:b3}), while in the HBA it scales as $V^2$ (Eq.\ref{eq:10}). Similarly, the third symplectic eigenvalue scales as $V$ in the CMA (Eq.\eqref{eq:b4}) and as $V^{1/2}$ in the HBA (Eq.\eqref{eq:14}). Consequently, the Holevo bound grows significantly faster in CMA than in HBA, preventing the system from reaching an asymptotic SKR for large $V$ due to the rapid increase of the Holevo bound (as reference see Fig.\ref{fig:05}).\\

\begin{figure}[h]
   \centering
    \begin{subfigure}[b]{0.48\textwidth}
         \centering
         \includegraphics[width=\textwidth]{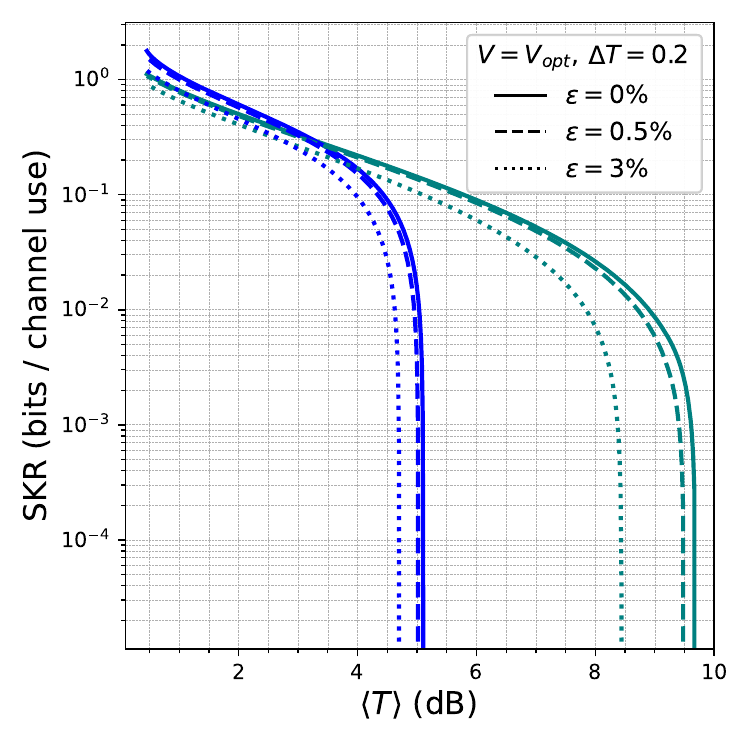}
     \end{subfigure}
         \centering
   \begin{subfigure}[b]{0.48\textwidth}
         \centering
         \includegraphics[width=\textwidth]{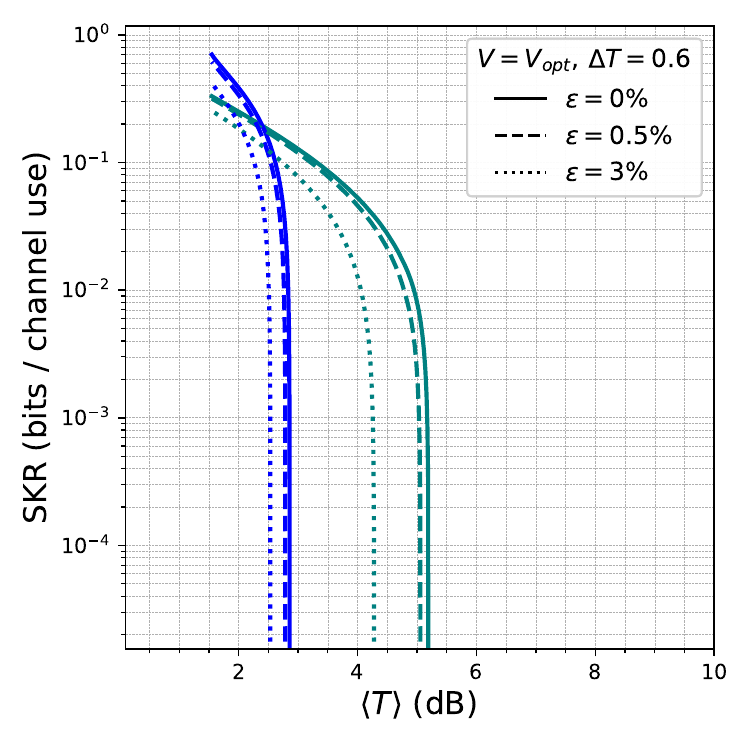}
     \end{subfigure}
 \caption{This Figure presents the key distribution as a function of the average transmittance $\avg{T}$, at the optimal variance for the CMA approach (green lines) and at a fixed variance of $V=10^3$ for the HBA (blue lines). The widths of the probability distribution are $\Delta T=0.2$ (left) and $\Delta T=0.6$ (right).}
\label{fig:04}
\end{figure}

For this reason, in the CMA it is necessary to optimize the variance as a function of the channel statistics, $V_{\text{opt}}(T_{\text{min}}) \equiv \max_V[R(V;T_{min})]$  giving the results shown in Fig.~\ref{fig:04}.
Indeed, it is fair to assume that the legitimate parties know the statistical properties of the channel and work at the optimal modulation variance.
In contrast, for the CMA, the optimal configuration naturally tends toward the high-variance limit, where the SKR becomes less sensitive to further increases in $V$.\\


\section{Conclusion}

This paper compares two different approaches for the evaluation of the secret key rate of CV-QKD on a fluctuating channel found in the literature.

The HBA approach is characterized by a slower growth of the Holevo bound with respect to the variance $V$, which allows the system to reach a stable asymptotic SKR for $V \gg 1$.
The choice of a rate matching the minimum transmittance of the channel represents a strong limitation for high attenuation channels, only partly mitigated by the lower estimation of the Holevo bound.
    
The CMA approach, on the other hand, is characterized by the choice of an optimized rate between the legitimate parties, which justifies the improvement in the high attenuation regime.
However, the increased excess noise introduced by the average of the covariance matrix is detrimental for a high level of fluctuations, requiring also a fine tuning of the modulation variance in order to get a positive key rate.

This work has studied through numerical simulations the difference in the obtained SKR for the two approaches.
The CMA approach seems more in line with what is done in experimental CV-QKD, where the Holevo bound is calculated from the observed statistics of the state after passing through the channel.
However, the simulations have shown that the HBA tends to underestimate the SKR for all the configurations examined and its not requiring an optimization on the modulation variance still makes it a valuable alternative for the estimation of the performance of CV-QKD on the fluctuating channel.

In addition to this, the effects of the choice of a lower code rate in the HBA approach and its impact on the leaked information need to be further investigated.

\section*{Acknowledgments}
This research was supported by the Q-CAYLE project, funded by the European Union-Next Generation UE/MCIU/Plan de Recuperacion, Transformacion y Resiliencia/Junta de Castilla y Leon (PRTRC17.11), and also by RED2022-134301-T and PID2023-148409NB-I00, financed by MI-CIU/AEI/10.13039/501100011033.  
The financial support of the Department of Education of the Junta de Castilla y León and FEDER Funds is also gratefully acknowledged (Reference: CLU-2023-1-05).
MS acknowledges funding from the European Union’s Horizon 2020 research and innovation
programme under grant agreement No 101082596 (project QUDICE).

\appendix
\section{Asymptotic approximation}\label{appendix0}
In this section, the eigenvalues used in Section~\ref{large_variance} are developed in the limit $V 
\gg 1$. In this way, Eq.~\eqref{eq:10} in terms of the noise variance $\omega$ are given by 
\begin{align*}
    A=&V^2(1-2T)+2T+T^2\left(V+\frac{(1-T)\omega}{T}\right)^2,\\
    &B=T^2\left(V\frac{(1-T)\omega}{T}+1\right)^2.
\end{align*}
Taking into account the dominant terms of $V$
\begin{equation}
    A=V^2(1-2T)+T^2V^2=V^2(1-T)^2,\quad B=V^2(1-T)^2\omega^2.
\end{equation}
Therefore, the expression for the eigenvalues has the form 
\begin{equation}
    \lambda_{1,2}=\frac{1}{\sqrt{2}}\left(V^2(1-T)^2\pm V^2(1-T)^2\sqrt{1-\frac{4\omega^2}{V^2(1-T)^2}}\right)^{1/2}.
\end{equation}
Using the approximation $\sqrt{1\pm a\,x^2}\approx 1\pm a\,x^2/2$ for $x\ll1$ is used, where $a\rightarrow 4\omega^2/(1-T)^2$ and $x\rightarrow 1/V$, and substituting into the above equation, the result is obtained
\begin{equation}
    \lambda_{1,2}\approx\frac{1}{\sqrt{2}}\left(V^2(1-T)^2\pm V^2(1-T)^2\left(1-\frac{2\omega^2}{V^2(1-T)^2}\right)\right)^{1/2}.
\end{equation}
Considering that $V\gg 1$ again, the eigenvalues $\lambda_{1,2}$ are given by
\begin{align}
    \lambda_{1}=&\frac{1}{\sqrt{2}}\left(2\,V^2(1-T)^2-2\,\omega^2\right)^{1/2}\approx V\,(1-T),\nonumber\\
    &\lambda_2=\frac{1}{\sqrt{2}}(2\,\omega^2)^{1/2}=\omega.
\end{align}
Based on the preceding development, the third eigenvalue for $V\gg1$ is expressed as
\begin{equation*}
    \lambda_3\approx\sqrt{\frac{(1-T)\,\omega\, V}{T}}.
\end{equation*}
\section{Holevo bound asymptotically function}\label{appendix1}
This appendix presents the derivation of the Holevo bound corresponding to the HBA approach. Within this framework, the Holevo bound is expressed as the mean value of the $\chi_{BE}$ function
\begin{equation*}
    \avg{I_{BE}}=\frac{1}{\Delta T}\int_{T_{min}}^{T_{min}+\Delta T}\chi_{BE}\,\,dT,
\end{equation*}
the indefinite integral has the following form 
\begin{equation*}
   \avg{\chi_{BE}}=\frac{1}{2\,\Delta T} \left(\frac{\log_2 (\overline{T}+\varepsilon\,T)}{1-\varepsilon}+T \left(\log_2 \left(\frac{\overline{T}^2 T\,V}{\overline{T}+\varepsilon\,T}\right)-2\log_2(e)\right)- \log_2 \overline{T}^2\right)+\Big\langle h\left(\frac{\omega-1}{2}\right)\Big\rangle,
\end{equation*}
where the function $h$ is defined as
\begin{equation*}
     h\left(\frac{\omega-1}{2}\right)=\left(\frac{\omega+1}{2}\right)\log_2\left(\frac{\omega+1}{2}\right)-\left(\frac{\omega-1}{2}\right)\log_2\left(\frac{\omega-1}{2}\right),
\end{equation*}
and $\omega$ denotes the thermal variance introduced in Eq.~\eqref{eq:18}. Consequently, the mean value of this term can be written as
\begin{align}
    \Big\langle h\left(\frac{\omega-1}{2}\right)\Big\rangle=\frac{1}{2\Delta T} \bigg[&  2 \log _2\overline{T}-\frac{2 (\varepsilon -1)}{\varepsilon -2}\log_2 T+\frac{2}{\varepsilon -2}\log _2\left(2\overline{T}+\varepsilon\,T\right)+T \log _2\left(\frac{\varepsilon\,T}{4 \overline{T}^2}(2\overline{T}+\varepsilon\,T)\right)\nonumber\\
    &-\frac{\varepsilon -1}{\varepsilon -2}(2 \overline{T}+\varepsilon\,T)\log _2\left(\frac{2\overline{T}+\varepsilon\,T}{\varepsilon\, T}\right)
     -\frac{\varepsilon}{\ln (2)}\left(\text{Li}_2(\overline{T})-\text{Li}_2\left(\frac{(\varepsilon -2)\overline{T}}{\varepsilon }\right)\right)\bigg].
     \label{Appendix_B3}
\end{align}
The evaluation of $\avg{\chi_{BE}}$ over the transmittance interval yields the Holevo bound for the HBA approach
\begin{equation*}
    \avg{I_{BE}}=\avg{\chi_{BE}}|_{\Delta T+T_{min}}-\avg{\chi_{BE}}|_{T_{min}}.
\end{equation*}
Finally, the auxiliary function $\widetilde{h}$, which completes the integral, corresponds to the evaluation of Eq.~\eqref{Appendix_B3} at the extremes of the transmittance interval. Its explicit form is given by
\begin{align*}
    \widetilde{h}&=\frac{1}{2\Delta T} \bigg[  2 \log _2\left(\frac{\overline{T}_{max}}{\overline{T}_{min}}\right)-\frac{2 (\varepsilon -1)}{\varepsilon -2}\log_2\left(\frac{T_{max}}{T_{min}}\right)+\frac{2}{\varepsilon -2}\log _2\left(\frac{2\overline{T}_{max}+\varepsilon\,T_{max} }{2\overline{T}_{min}+\varepsilon\,T_{min}}\right)\nonumber\\
    &+T_{max} \log _2\left(\frac{\varepsilon\,T_{max}}{4 \overline{T}_{max}^2}(2\overline{T}_{max}+\varepsilon\,T_{max})\right)-\frac{\varepsilon -1}{\varepsilon -2}(2 \overline{T}_{max}+\varepsilon\,T_{max})\log _2\left(\frac{2\overline{T}_{max}+\varepsilon\,T_{max}}{\varepsilon\, T_{max}}\right)\nonumber\\
   &-T_{min} \log _2\left(\frac{\varepsilon\,T_{min}}{4\overline{T}_{min}^2}(2\overline{T}_{min}+\varepsilon\,T_{min})\right)+\frac{\varepsilon -1}{\varepsilon -2}(2\overline{T}_{min}+\varepsilon\,T_{min}) \log_2\left(\frac{2\overline{T}_{min}+\varepsilon\,T_{min}}{\varepsilon\,T_{min}}\right)\nonumber\\
     &-\frac{\varepsilon}{\ln (2)}\left(\text{Li}_2(\overline{T}_{max})-\text{Li}_2\left(\frac{(\varepsilon -2)\overline{T}_{max}}{\varepsilon }\right)-\text{Li}_2(\overline{T}_{min})+\text{Li}_2\left(\frac{(\varepsilon -2)\overline{T}_{min}}{\varepsilon }\right)\right)\bigg],
\end{align*}
in this case, the dilogarithm function $\text{Li}_2(z)=\int_z^0 \ln(1-t)/t\,dt$. In the limit of low excess noise $\varepsilon\rightarrow 0$, or equivalently $\omega\rightarrow 1$, the $h$ function vanishes,
\begin{equation}
    \lim_{\varepsilon\rightarrow 0}h\left(\frac{\omega-1}{2}\right)=0,
\end{equation}
which corresponds to the case of a passive eavesdropper.

\section{Holevo bound in the CMA approach}\label{appendix2}
In the CMA approach, the Holevo bound~\eqref{eq:34} depends to the symplectic eigenvalues of the covariance matrix~\eqref{eq:27}, which are given by
\begin{align}
    &\widetilde{\lambda}_{1,2}=\sqrt{\frac{1}{2}}\sqrt{A\pm\sqrt{A^2-4B}},\nonumber\\
    A=V^2(1-2\,T_{eff}&)+2\,T_{eff}+T_{eff}^2(V+\chi_{eff})^2,\quad 
    B=T_{eff}^2\left(V\,\chi_{eff}+1 \right)^2,\nonumber\\
    &\widetilde{\lambda}_3=\sqrt{V \left(\frac{1+V\chi_{eff}}{V+ \chi_{eff}}\right)}.
\end{align}
In this manner, if the highest power of the variance is factorized on the A and B terms, the following expressions are obtained
\begin{align}
    &A=\left((1-2T_{eff})+2\frac{T_{eff}}{V^2}+A_0^2\right)\,V^2, \quad A_0=T_{eff}\left(1+a+\frac{b}{V}\right),\nonumber\\
    &B=T_{eff}^2\left(a(T_{eff})+\frac{b(T_{eff})}{V}+\frac{1}{V^2}\right)^2V^4\equiv B_0^2\,V^4.
    \label{eq:b3}
\end{align}
where $a$, $b$ are given by
\begin{align*}
    a(T_{eff})=\frac{\text{Var}(\sqrt{T})}{T_{eff}}+\varepsilon\left(1+\frac{\text{Var}(T)}{T_{eff}}\right),\qquad b(T_{eff})=\frac{1-\langle T\rangle}{T_{eff}}.
\end{align*}
The factorization of $V^2$ into the expression for $B$ is possible only because of the linear dependence of the effective noise $\chi_{eff}$ of variance $V$ in Eq.~\eqref{eq:30},
\begin{equation*}
    \chi_{eff}=a(T_{eff})V+b(T_{eff}).
\end{equation*}
Considering the above, together with the symplectic eigenvalues, the resulting expressions can be written as follows
\begin{align}
    \widetilde{\lambda}_{1,2}=&\left(\sqrt{\frac{1}{2}}\sqrt{A_0\pm\sqrt{A_0^2-4B_0^2}}\right)\,V\nonumber\\
    &\widetilde{\lambda}_3=\sqrt{\frac{B_0}{A_0}}V.
    \label{eq:b4}
\end{align}

\bibliography{sample}
\end{document}